\documentclass[aps,prb,twocolumn,superscriptaddress,longbibliography,noeprint]{revtex4-2}

\usepackage{amsmath}
\usepackage{graphicx}
\usepackage{bm}
\usepackage{multirow}
\usepackage[table]{xcolor}
\usepackage[export]{adjustbox}
\makeatletter
\let\old@makecaption=\@makecaption
\usepackage{subcaption}
\let\@makecaption=\old@makecaption
\makeatother
\usepackage{color}
\usepackage{orcidlink}

 \usepackage[normalem]{ulem}
 \usepackage{color}
  \usepackage{xspace}
 \usepackage{natbib}

\newcommand{\arc}{$\alpha$-RuCl$_{3}$}
\newcommand{\nio}{Na$_2$IrO$_{3}$}

\newcommand{\bs}[1]{\boldsymbol{#1}}

\newcommand{\pred}[1]{{\textcolor{black}{#1}}}

\begin{document}

	\title{Magnetic excitations in the Kitaev material Na$_2$IrO$_3$ studied by neutron scattering}
	
	\author{Alexandre Bertin\,\orcidlink{0000-0001-5789-3178}}
	\affiliation{Institute of Physics II, University of Cologne, 50937 Cologne, Germany}	
	\author{Hengdi Zhao\,\orcidlink{0000-0003-4606-6323}}
    \affiliation{Department of Physics, University of Colorado, Boulder, CO 80309, USA}		
	\author{Gang Cao\,\orcidlink{0000-0001-9779-430X}}
	\affiliation{Department of Physics, University of Colorado, Boulder, CO 80309, USA}		
	\author{Andrea Piovano\,\orcidlink{0000-0002-5005-6307}}
    \affiliation{Institut Laue-Langevin, 71 Avenue des Martyrs, Grenoble, 38042, cedex 9, France}	
	\author{Paul Steffens\,\orcidlink{0000-0002-7034-4031}}
    \affiliation{Institut Laue-Langevin, 71 Avenue des Martyrs, Grenoble, 38042, cedex 9, France}
    \author{Alexandre Ivanov\,\orcidlink{0000-0002-5092-8937}}
    \affiliation{Institut Laue-Langevin, 71 Avenue des Martyrs, Grenoble, 38042, cedex 9, France}
	\author{Markus Braden\,\orcidlink{0000-0002-9284-6585}}	
	\affiliation{Institute of Physics II, University of Cologne, 50937 Cologne, Germany}

	\begin{abstract}	
		
Inelastic neutron scattering experiments with a large set of  comounted Na$_2$IrO$_3$ crystals reveal the low-energy magnon dispersion in this candidate material for Kitaev physics. The magnon gap amounts to 1.7(1)\,meV and can be interpreted similarly to the sister compound $\alpha$-RuCl$_{3}$ to stem from the \pred{zone boundaries in the antiferromagnetic zigzag structure}. The neutron experiments find no evidence for low-energy excitations with ferromagnetic character, which contrasts to the findings in \arc .
\pred{Our results are consistent with a recently proposed microscopic model that involves an antiferromagnetic Heisenberg nearest-neighbor exchange in Na$_2$IrO$_3$ in contrast to the ferromagnetic one considered for $\alpha$-RuCl$_{3}$.
Although the magnetic response shows the signatures of bond-directional anisotropy in both materials the different relative signs of Kitaev and Heisenberg interaction result in different deviations from the initial Kitaev model.} Low-energy ferromagnetic fluctuations cannot be considered as a fingerprint of ferromagnetic Kitaev interaction.   
	
	\end{abstract}
	
	\date{\today}
	
	\maketitle

\section{Introduction}

Systems with bond-dependent exchange couplings, i.e.\ compass models, have attracted a lot of attention over  the past decade~\cite{Nussinov15}. One of the most intriguing system is the so-called Kitaev model consisting of transition-metal ($TM$) spin-1/2 arranged on a honeycomb lattice. Its exact solution predicts a quantum spin liquid (QSL) phase supporting anyon Majorana excitations~\cite{Kitaev06} \pred{ that are relevant for quantum computing}~\cite{Nayak08}. 
The peculiar geometry with  oxygen surrounding in form of
edge-sharing octahedra enforces symmetric 90$^{\circ}$ $TM$-O-$TM$ bonds and result\pred{s} in reduction of the Heisenberg superexchange interaction~\cite{Jackeli09,Chaloupka10}. In addition, strong spin-orbit coupling (SOC) acting on a low-spin-state $d^5$ electronic configuration in an undistorted octahedral environment implies an effective spin $j_{\rm eff}=\frac{1}{2}$ ground state~\cite{Jackeli09}. The bond-directional character of the magnetic interaction results from the Ising coupling of different effective-spin components for different nearest-neighbor $TM$-$TM$ bond orientation.

The quest for materials fulfilling the Kitaev criteria started with a scarce number of candidates: Na$_2$IrO$_3$~\cite{Singh10}, $\alpha$-Li$_2$IrO$_3$~\cite{Singh12}, or $\alpha$-RuCl$_3$~\cite{Plumb14}. A few other 2D materials with chemical formula $A_3$LiIr$_2$O$_6$, extension to 3D hyperhoneycomb lattice ($\beta$-Li$_2$IrO$_3$, $\gamma$-Li$_2$IrO$_3$), but also $3d$ $TM$ and $4f$ rare-earth  compounds have garnished the realm of the \pred{Kitaev physics~\cite{WitczakKrempa14,Schaffer16,Rau16,Cao18,Winter17,Takagi19,Liu20,Motome20,Trebst22,Rousochatzakis24,Jang24}.}
 
For \pred{four-valent Ir in an undistorted cubic ligand field, the $t_{2g}$ states} are split because of the strong SOC into a lower filled $j_{\rm eff}=\frac{3}{2}$ and an upper half-filled $j_{\rm eff}=\frac{1}{2}$ band. The combined effect of SOC and Coulomb interaction further opens a gap in the $j_{\rm eff}=\frac{1}{2}$ band into upper and lower Hubbard bands, yielding a Mott insulating state, as it is observed in Sr$_2$IrO$_4$~\cite{Kim08,Kim09} and in related Ruddlesden-Popper iridates~\cite{Rau16,Cao18}. Na$_2$IrO$_3$ falls also into the category of the Mott-type topological insulator, as revealed by angle-resolved photoemission where the insulating gap $\Delta \approx 340$\,meV is found temperature independent~\cite{Comin12}, and it is also coined a quantum spin Hall insulator due to the prediction of helical edge states~\cite{Shitade09}.

\pred{Na$_2$IrO$_3$ crystallizes in the $C2/m$ space group with lattice parameters $a=5.43$\,\AA \ perpendicular to the Ir-Ir bonds, $b=9.40$\,\AA \ parallel to the bonds, and $c=5.61$\,\AA, $\beta_m=109.04^{\circ}$~\cite{Choi12,Ye12}. 
All reciprocal-space vectors  are given in reduced lattice units (r.l.u.) with respect to this lattice.
Neighboring layers are shifted parallel to the
$\boldsymbol a$ axis by about one third of the $a$ lattice constant so that the third-nearest layer lies almost
perfectly above the initial one. Note that the reciprocal lattice vector $\boldsymbol b^\star$ is parallel to $\boldsymbol b$ and has the length of $\frac{2\pi}{b}$, but $\boldsymbol c^\star$ is not parallel to $\boldsymbol c$.  $\boldsymbol c^\star$ is perpendicular to the
layers and has the length $\frac{2\pi}{d}$ with $d$ the inter-layer distance of $\sim$5.3\,\AA. Our sample unit shows domains rotated by $\pm$120\,$^\circ$ around $\boldsymbol c^\star$ following the $C_3$ symmetry of a single IrO$_3$ layer. The domain distribution is even more complex when taking the 3D stacking into account~\cite{Braden2025}, but throughout this work we only discuss the 2D aspects of magnetic excitations.
}

\pred{Na$_2$IrO$_3$ undergoes a magnetic transition towards AFM order at $T_{\rm N} = 15$\,K, with a magnetic propagation wavevector $\boldsymbol q_{\rm mag}=(0,\,1,\,0.5)$. 
Note that this ordering breaks the translation symmetry of a single honeycomb layer.
In the zigzag-type magnetic order~\cite{Ye12,Choi12} moments are aligned antiferromagnetically along an Ir-Ir bond (Z bond parallel to the ${\bm b}$ axis), and ferromagnetically aligned in the direction perpendicular to the Z bond (along the ${\bm a}$ axis). Two neighboring honeycomb layers are coupled antiferromagnetically. Neutron diffraction finds that the magnetic moment mostly aligns along the ${\bm a}$ axis~\cite{Liu11} and amounts to $\sim$0.2~$\mu_{\rm B}$~\cite{Ye12}. In contrast, diffuse magnetic x-ray experiments reported that the moments are canted by an angle of $\approx 44^{\circ}$ towards the vertical axis~\cite{Chun15}.}

Specific-heat and susceptibility measurements on Na$_2$IrO$_3$ \pred{show the 
magnetic transition at $T_{\rm N}=15$\,K~\cite{Singh10,Takagi19}, and a negative Curie-Weiss temperature suggests dominant antiferromagnetic (AFM) interaction and a moderate frustration ratio.}
%Neutron diffraction studies have shown that the magnetic order is of the zigzag type, with magnetic propagation wave vector ${\boldsymbol q}_{\rm mag}=(0,\,1,\,0.5)$~\cite{Ye12,Choi12}. 
Despite the occurrence of magnetic ordering, this material could lie near the Kitaev QSL phase. 
Specific-heat \pred{data} yielded support for the proximity of the QSL phase~\cite{Mehlawat17}, as a broad hump at high temperature in the magnetic specific heat was interpreted in terms of spin fractionalization. 
Concerning the magnetic excitation spectrum, resonant inelastic x-ray spectroscopy (RIXS) observed a broad continuum persisting up to room temperature~\cite{Gretarsson13a,Revelli20}, which is reminiscent of the low-energy continuum found in $\alpha-$RuCl$_3$~\cite{Banerjee17}. 
The corresponding dynamical structure factor was interpreted by spin-spin correlations restricted to the Kitaev interaction on the nearest-neighbors~\cite{Revelli20}, while 
RIXS studies with higher resolution revealed a clear dispersion that gives access to the microscopic interaction parameters~\cite{Kim20}. 
The bond-directional nature of the magnetic couplings has been revealed by resonant elastic x-ray measurements that probe the diffuse magnetic correlations~\cite{Chun15} and more recently by a scattering-vector analysis
of the high-energy RIXS response~\cite{Magnaterra23}.

\pred{ \arc ~ is a $4d$ ruthenium-based candidate for Kitaev physics with similar honeycomb layers. Also this material exhibits the nearly identical antiferromagnetic zigzag magnetic structure documenting deviations from the ideal Kitaev model \cite{Plumb14,Takagi19,Trebst22}.
Compared to \nio ,   \arc \ offers better conditions for inelastic neutron scattering (INS) experiments that are almost completely missing for \nio . Several INS studies on \arc \  reported the dispersion of magnetic excitations  \cite{Banerjee16,Banerjee17,Banerjee18,Balz19,ran2017,ran2022,do2017,Braden2025} yielding an
estimation of the relevant interaction parameters (see Ref. \cite{moeller2025} for a recent summary). The magnetic excitations in \arc \ strongly deviate from the expectation for a simple isotropic antiferromagnet, in particular there are low-energy ferromagnetic (FM) fluctuations indicating competition of FM and AFM ordering schemes, and at higher temperatures the excitations with FM character even dominate \cite{Banerjee16,Banerjee17,Banerjee18,Balz19,ran2017,ran2022,do2017,Braden2025}.}

Over the past decade, in order to explain the magnetic order and the dispersion of magnetic excitations, the Kitaev Hamiltonian was considerably extended by Heisenberg interactions up to the 3rd-nearest neighbors,
$J_1$, $J_2$ and $J_3$, and by further bond-directional symmetric off-diagonal terms called $\Gamma$ and $\Gamma'$ \cite{Foyevtsova13,Katukuri14,Chaloupka10,Chaloupka13,Foyevtsova13,Chaloupka15,Winter17,moeller2025}. The commonly accepted minimal $HK\Gamma\Gamma'$ Hamiltonian with the nearest-neighbor anisotropic interaction corresponding to the bond type $\zeta=X,Y,Z$ is:
\begin{align}
	\begin{split}	
		\mathcal{H}^{\zeta}=&\sum_{ \langle i,j \rangle } \Bigl\{ J_1 \bm{S}_i\bm{S}_j+KS_i^{\alpha} S_j^{\alpha} +\Gamma (S_i^{\beta} S_j^{\gamma} +S_i^{\gamma} S_j^{\beta})\\
		&+\Gamma^{\prime}(S_i^{\alpha} S_j^{\beta} +S_i^{\beta} S_j^{\alpha}+S_i^{\alpha} S_j^{\gamma} +S_i^{\gamma} S_j^{\alpha}) \Bigr\} \\
		&+J_2\sum_{\langle\langle i,j \rangle\rangle} \bm{S}_i\bm{S}_j+J_3 \sum_{\langle\langle\langle i,j \rangle\rangle\rangle} \bm{S}_i\bm{S}_j, 
	\end{split}	
	\label{hamiltonian_Kitaev}           
\end{align}
where $(\alpha,\beta,\gamma)$ are the cubic spin coordinates $(x,y,z)$, $(y,z,x)$, and $(z,x,y)$ for bonds $\zeta=X,Y,Z$, respectively. The bond-directional parameters are $K$ the Kitaev exchange, $\Gamma$ and $\Gamma^{\prime}$ the symmetric off-diagonal exchange terms, and $J_{1,2,3}$ denote the Heisenberg exchange between nearest-neighbors ($\langle ... \rangle$), 2nd nearest-neighbors ($\langle\langle... \rangle\rangle$), and 3rd nearest-neighbors ($\langle\langle\langle... \rangle\rangle\rangle$), respectively. While there is no doubt that \nio \ exhibits a strong Kitaev interaction, $K$, it remains an open issue whether it is dominant enough to imply Kitaev physics such as the QSL state.

Here, we report on INS studies with two triple-axis spectrometers (TAS) that could overcome
the absorption and sample-size issues of \nio \ and that reveal the low-energy part of the spin-wave dispersion
in Na$_2$IrO$_3$. \pred{Thereby, we complement the high-energy RIXS data and we can evaluate various $HK\Gamma\Gamma'$ parameter sets of the magnetic interaction that were proposed in the literature.}

\begin{figure}[h!]
	\centering
	\includegraphics[width=0.77\columnwidth]{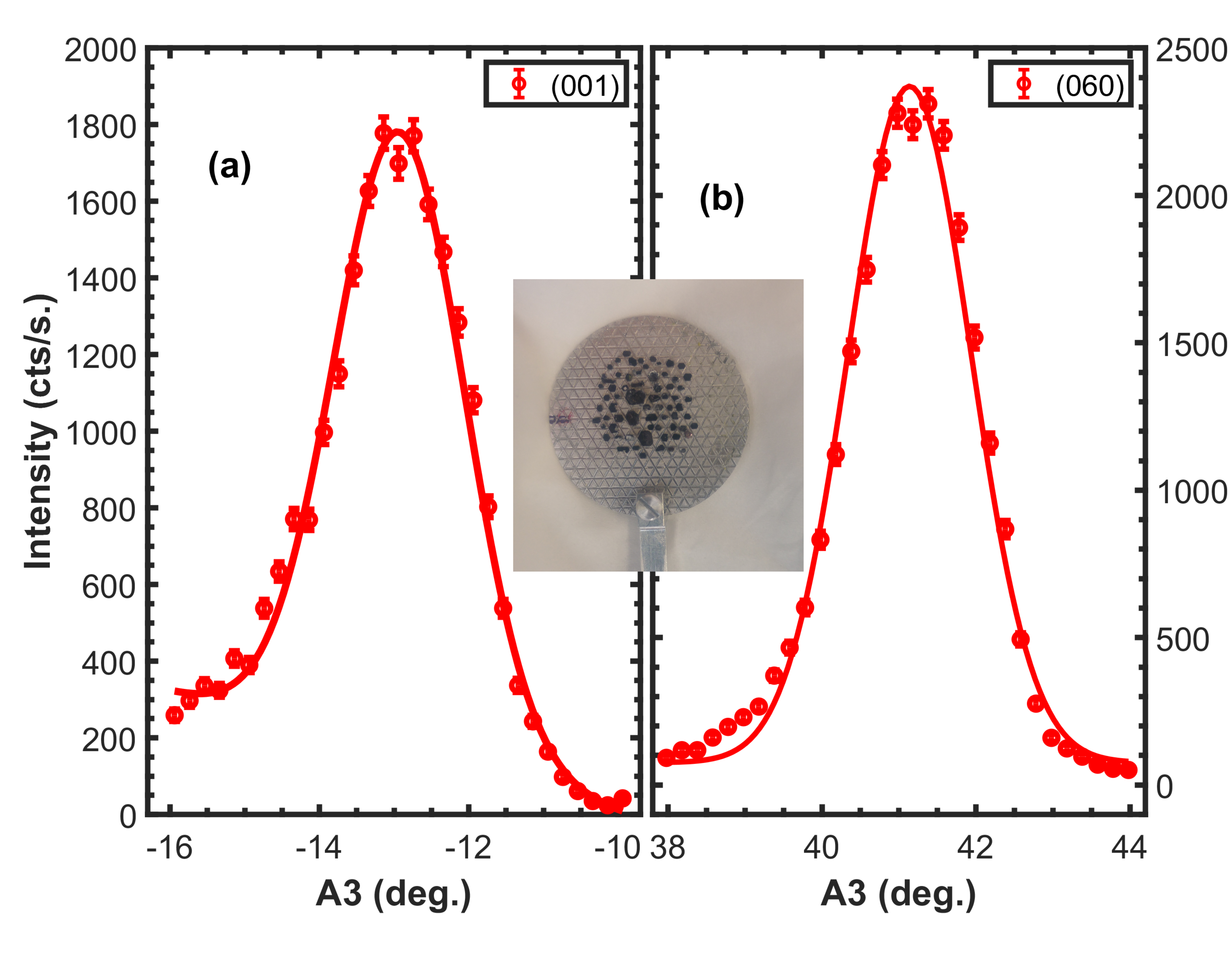}
	\caption{\label{sample_SM}  \pred{Rocking scans on the nuclear reflections  (0,\,0,\,1) (a) and (0,\,6,\,0) (b) measured on {\it IN8(2)}. The inset presents a photo of the 63 co-aligned Na$_2$IrO$_3$ single crystals glued on an  Aluminum plate with 30\,mm diameter.}}
\end{figure}

\section{Experimental}
\label{methods}

INS experiments were performed at the Institut Laue-Langevin (ILL, Grenoble) on the thermal neutron TAS {\it IN8}~\cite{refin8,data-in81,data-in82} and on the cold neutron TAS {\it Thales}~\cite{data-thales}, both equipped with focusing highly-oriented pyrolytic-graphite monochromator and analyzer units. \pred{For most scans the final neutron wavevector amounted to $k_{f}=2.662$\AA$^{-1}$~and 1.55\,\AA$^{-1}$ on {\it IN8} and {\it Thales}, respectively, resulting in an energy, $E$, resolution at the elastic line of 0.2\,meV and 0.8\,meV respectively.} A pyrolytic graphite filter and a Beryllium filter were used, respectively, in order to suppress higher order contamination. On {\it Thales} a neutron velocity selector was inserted in the incoming beam.
\pred{At both instruments the sample was cooled down to 1.5\,K in an ILL orange-type cryostat, and we performed most scans at this temperature.} Note that two different experiments have been carried out on {\it IN8} and are referred to in this manuscript as {\it IN8(1)}~\cite{data-in81} and {\it IN8(2)}~\cite{data-in82}. In the second  {\it IN8(2)} experiment the background could be considerably reduced by improving the positioning of 
diaphragms with respect to the outer walls of the cryostat.

\pred{\nio \ crystals} were grown using a self-flux method at University of Colorado, and similarly obtained samples were already studied in Ref. \cite{Ye12}. 
The crystals exhibit a plate shape with the large dimensions parallel to the
Ir layers and with a thickness of only some tenths of mm.
The crystalline quality of the crystals was analyzed by x-ray Laue diffraction.
In the inset of Fig.~\ref{sample_SM} we show the sample mounting used for the INS experiments. In order to cope with the strong neutron absorption of Iridium and to obtain sufficient sample volume, 63 thin single crystals have been co-aligned and mounted with an hydrogen-free glue on an Aluminum plate. In Fig.~\ref{sample_SM}(a) and (b) we show rocking scans, performed by rotating the A3 sample angle, on the (0,\,0,\,1) and (0,\,6,\,0) nuclear reflections, respectively. The full width at half maximum (FWHM) gives an estimation of the sample mosaicity $\eta \approx 2^{\circ}$, a reasonable value given the large number of co-aligned single crystals. Our sample shows the 120 degrees rotation of domains following the
$C_3$ symmetry of the single honeycomb layer.
In the scattering plane spanned by (0,\,1,\,0) and (0,\,0,\,1) we avoid that neutrons pass along the larger dimensions of the individual crystals
and of the total sample mounting. The absorption remains thus acceptable, but the main challenge arises from 
the still small size of the sample with a total mass of $\sim$208\,mg.

\pred{Spin wave calculations using linear spin-wave theory (LSWT) have been performed with the {\it SpinW} package~\cite{Toth15} implementing the magnetic-interaction model described in Equ.~(1).
}
 
 \section{Neutron Scattering Studies}
 \label{INS}

 \subsection{Magnetic Order}
 \label{INS_order}

 \begin{figure}[h!]
	\centering
	\includegraphics[width=0.50\columnwidth]{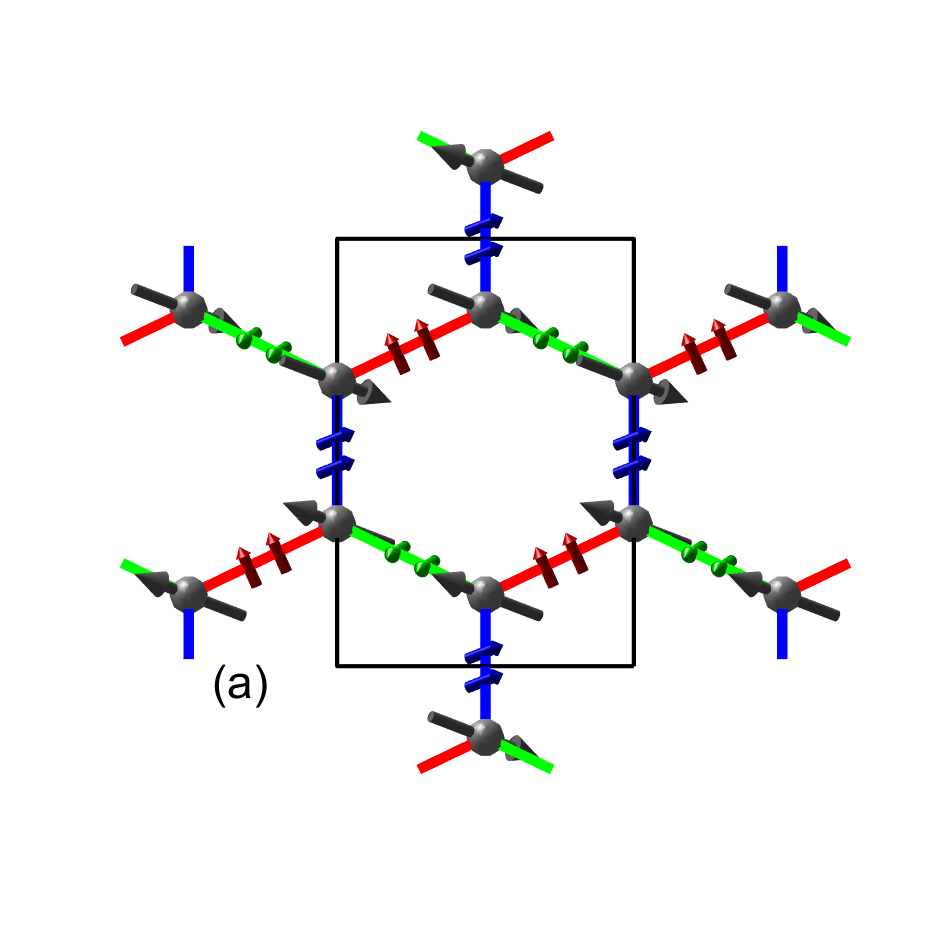}
	\includegraphics[width=1.00\columnwidth]{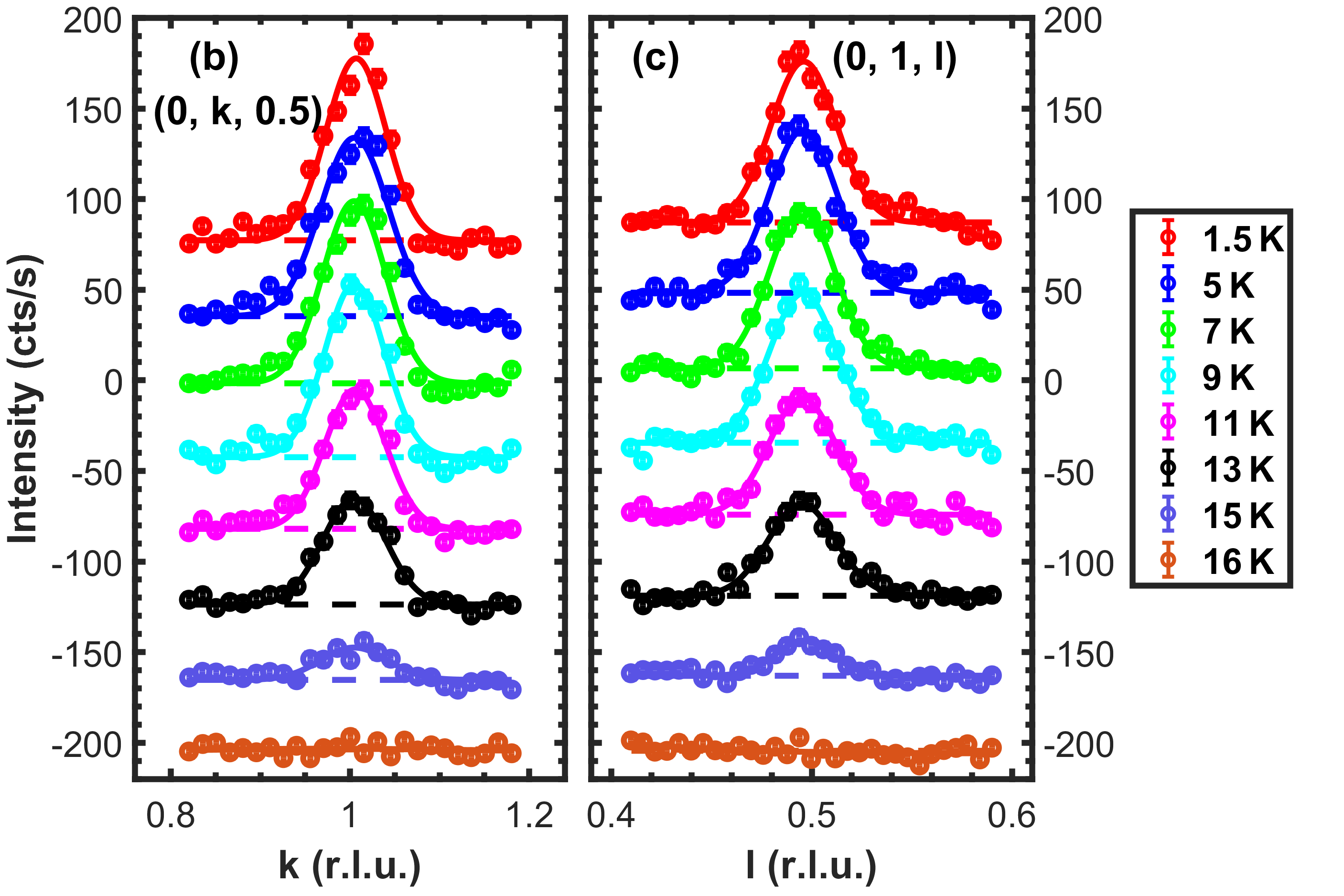}
	\includegraphics[width=1.00\columnwidth]{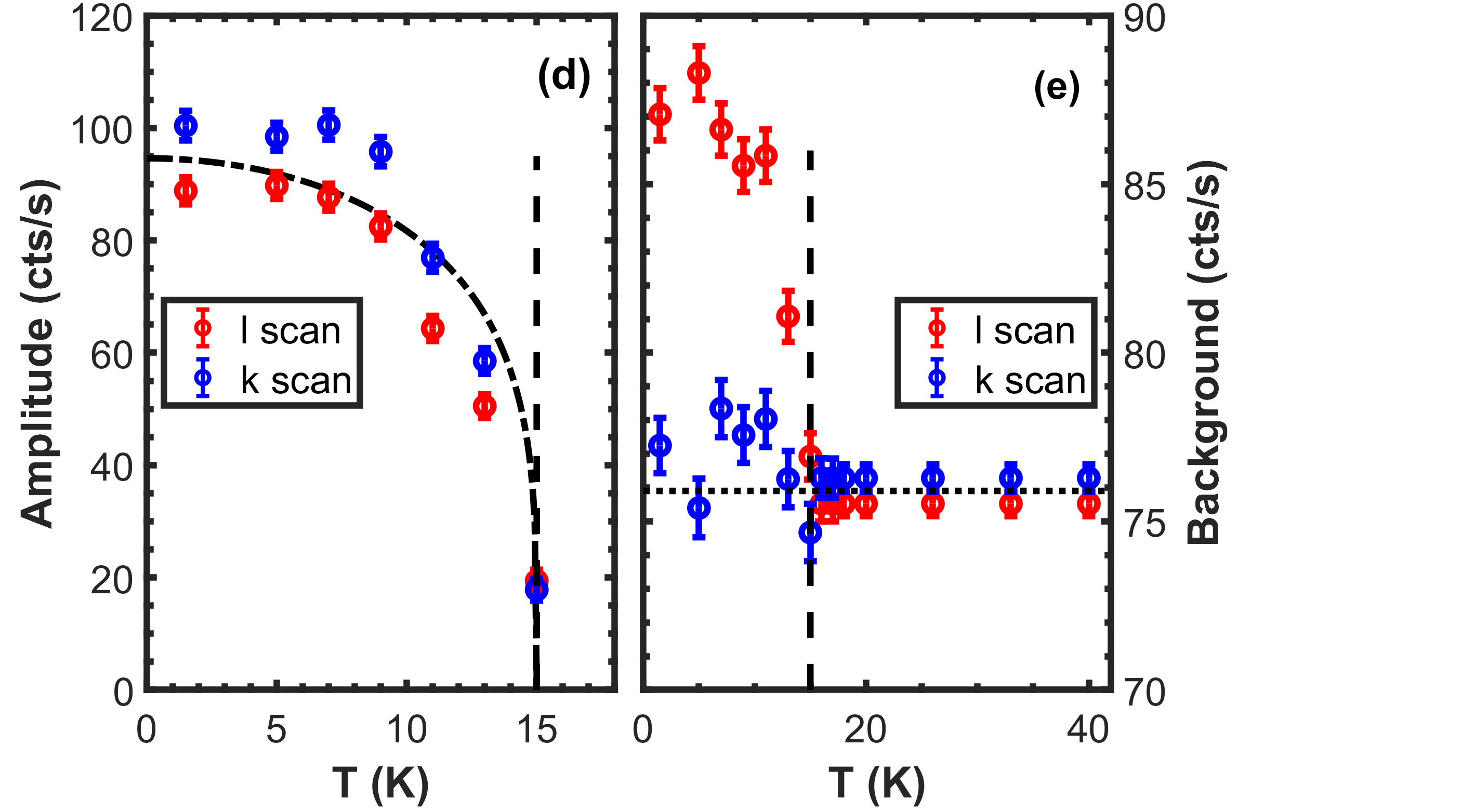}		
	\caption{\label{NIO_mBP} (a) Scheme of the zigzag magnetic order in \nio : Ir sites (spheres) and ordered moments (arrows) are indicated in gray for a single honeycomb layer. The red, green and blue thick lines indicate the $X$, $Y$ and $Z$ bonds, respectively, with the colored double arrows denoting the component ferromagnetically coupled by the Kitaev \pred{interaction} acting on this bond. The monoclinic in-plane lattice parameters are shown by the black rectangle with $\boldsymbol{a}$ perpendicular and  $\boldsymbol{b}$ parallel to an Ir-Ir bond.		
	\pred{On $Thales$, we scanned the (0,\,1,\,$\frac{1}{2}$) magnetic Bragg peak along the $k$ (b) and $l$ (c) directions; some data were shifted vertically for clarity \cite{vertical_shifts}.} Temperature dependence of the (0,\,1,\,$\frac{1}{2}$) magnetic Bragg peak amplitude (d) and background (e), for both scan directions. The vertical dashed line indicates $T_{\rm N}$. The dashed-dotted line in panel (d) is a guide to the eye, see main text. The horizontal dotted line in panel (e) indicates the paramagnetic background, averaged for both scan directions.}
\end{figure}

\pred{Figure 2(a) illustrates the magnetic order for a single honeycomb layer, and in Fig.~\ref{NIO_mBP}(b) and Fig.~\ref{NIO_mBP}(c),} we show elastic scans across the magnetic Bragg peak (0,\,1,\,0.5) taken along the in-plane $k$ and out-of-plane $l$ directions, respectively, on $Thales$. The Bragg peak intensities vanish above the N\'eel temperature $T_{\rm N}$ in agreement with previous studies~\cite{Choi12}. Since the magnetic Bragg peak develops also along the out-of-plane component, i.e.\ the $l$ direction, the order is threedimensional. \pred{The scan profiles are fitted with Gaussian functions,} and the temperature dependence of the magnetic Bragg-peak amplitudes for both scan directions, shown in Fig.~\ref{NIO_mBP}(d), confirms $T_{\rm N}$\,$\pred{\approx}$\,15\,K. The dashed-dotted black line is a guide to the eye, modeled with the phenomenological expression $y=y_0\cdot(1-(\frac{T}{T_{\rm N}})^2)^{2\beta}$, where the square exponent accounts for the quadratic decay at low temperature of the magnetic moment due to AFM magnons, and we chose $\beta=0.125$, the critical exponent of the order parameter for a 2D Ising system~\cite{leguillou1987,novotny1992}. \pred{Note, that the data density does not allow us free fitting of  $T_{\rm N}$ and $\beta$.} The FWHM is temperature independent and amounts to $\approx 0.08$ and 0.04 r.l.u. for the $k$ and $l$ direction, respectively, see Supplemental Material \cite{suppl-mat}. Finally, the temperature dependence of the background is shown in Fig.~\ref{NIO_mBP}(e). While for the $k$ direction the background remains unaffected by the magnetic transition, a clear uptake is visible at the transition for the scans along the $l$ direction indicative of short-range quasistatic magnetic correlations or magnetic stacking faults that emerge at the N\'eel temperature. This effect is sizable and indicates that analyzing the magnetic Bragg peaks will considerably underestimate the ordered moment. In contrast, we find no evidence for a persisting peak in the $k$ scan that would indicate strong 2D quasistatic correlations above $T_{\rm N}$.

\begin{figure}[t!]
	\centering
	\includegraphics[width=0.85\columnwidth]{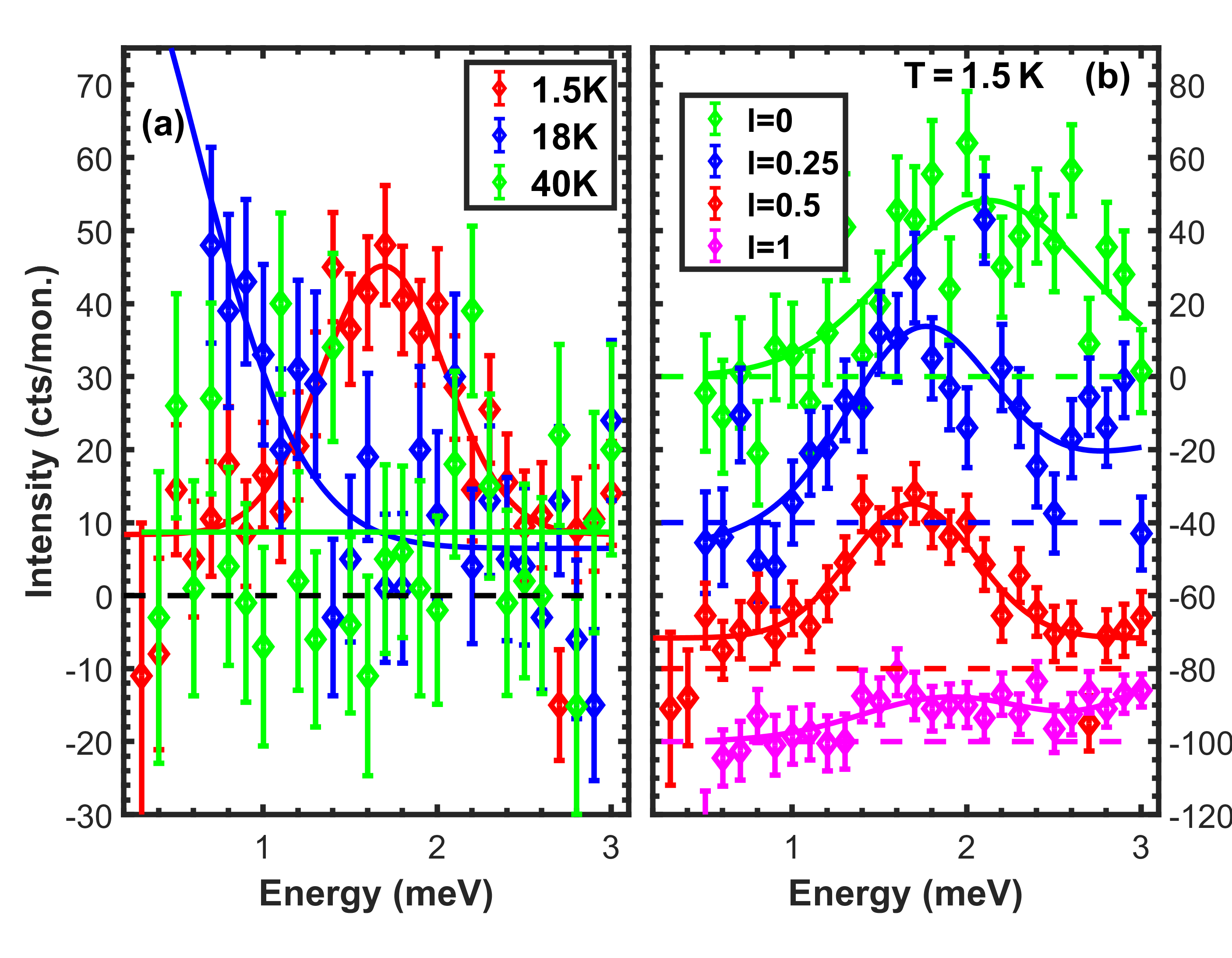}
	\includegraphics[width=0.85\columnwidth]{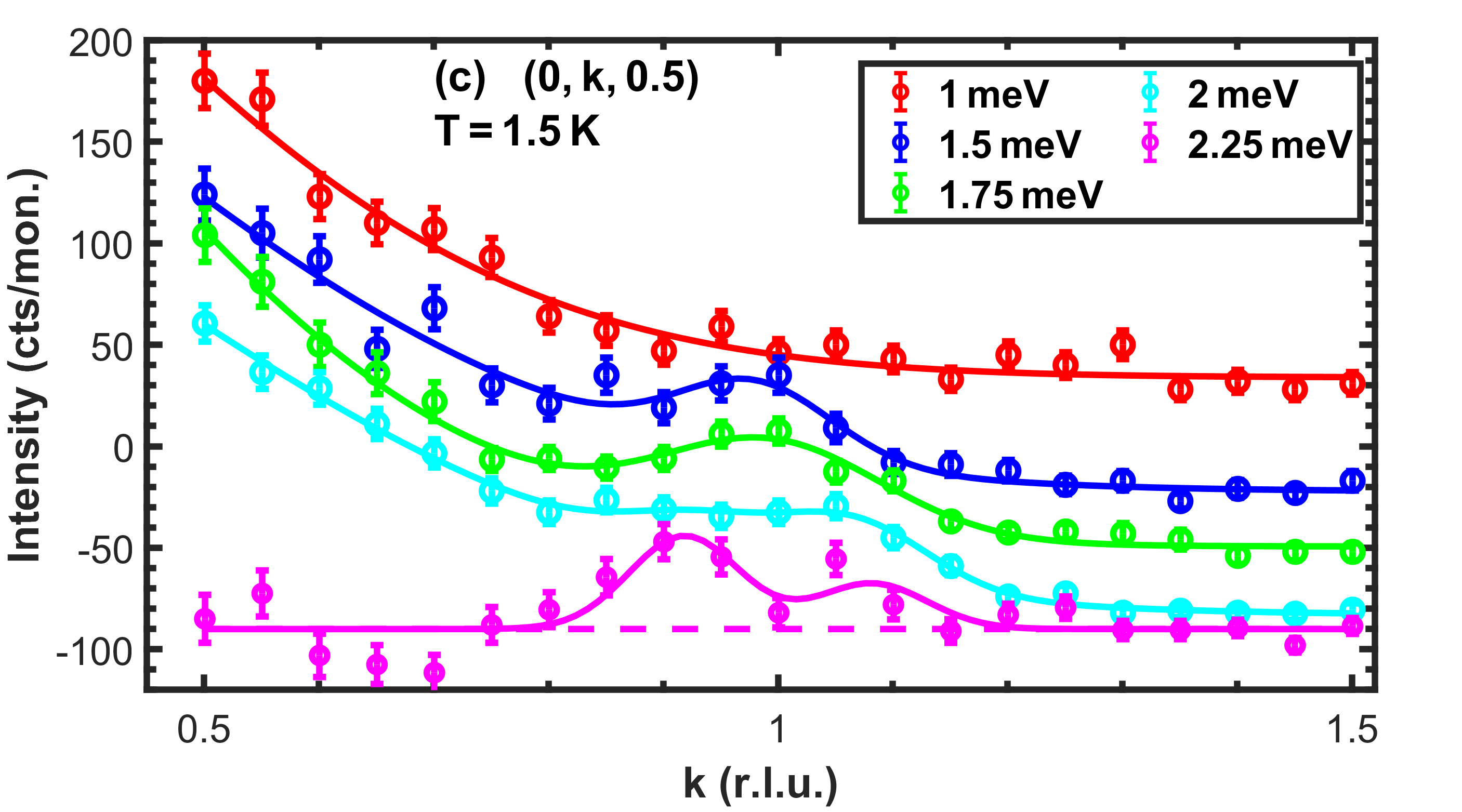}			
	\includegraphics[width=0.85\columnwidth]{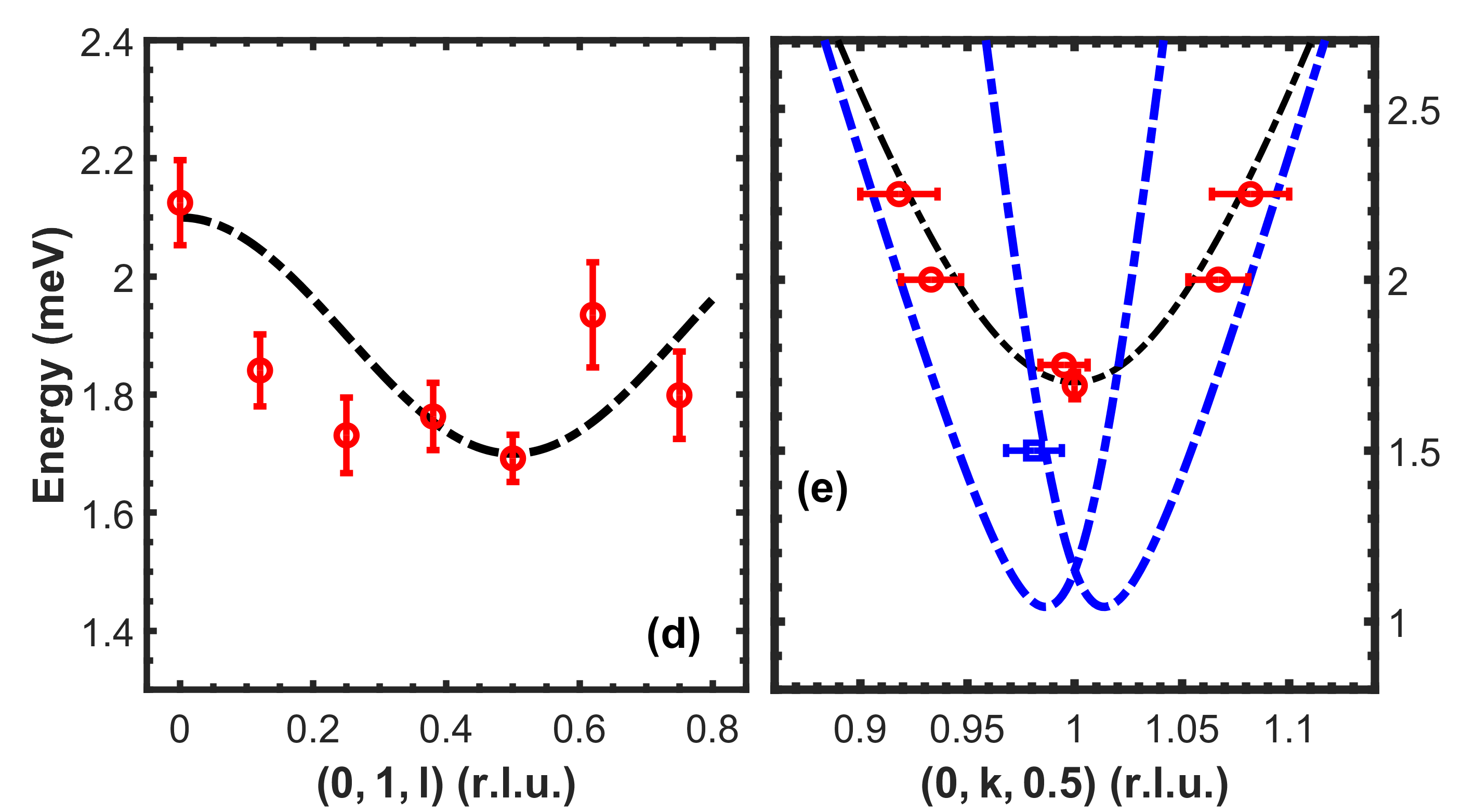}		
	\caption{\label{NIO_magnon_Thales} INS results obtained on the cold TAS $Thales$:  Constant-$\boldsymbol Q$ scans at the AFM zone center $(0, 1, 0.5)$ at various temperatures, (a), and at  $(0,\,1,\,l)$  for selected $l$ values at $T$\,=\,1.5\,K, (b). \pred{Background scans measured after rotating the sample by $\pm 45^{\circ}$ are subtracted from the data. (c) Constant-$E$ scans across the AFM zone center at $T$\,=\,1.5\,K. A background scan measured after rotating the sample by $45^{\circ}$  has been subtracted from the data collected at $E$\,=\,2.25\,meV, and the magenta dotted line indicates zero intensity. In panels (a-c), the monitor was set to $4\cdot10^6$~counts and solid lines are Gaussian fits to the data, see main text. From these fits the magnon dispersion along the $l$ (d) and the $k$ (e) directions is drawn. Black dashed-dotted lines in panels (d,e) are guides to the eye, while the blue dashed-dotted lines in (e) indicate the calculated spin-wave dispersion. The square blue symbol refers to the data point collected at $E=1.5$\,meV.} }
\end{figure}

\subsection{Magnetic Excitations}
\label{INS_excitations}

Despite being widely investigated, the microscopic parameters of the spin Hamiltonian remain elusive for \nio , because INS experiments remain scarce and only RIXS studies with an exceptional resolution can analyze the energy range of magnetic excitations in \nio~\cite{Kim20}.
Such experiments revealed magnetic excitations at the Brillouin zone center around 35\,meV and their downward
dispersion along paths into the Brillouin zone~\cite{Kim20}. 
However, even with the RIXS resolution of $\Delta E$\,=\,12\,meV it
is not possible to study lowest energies of the magnetic dispersion, i.e. well below 10\,meV.
Inelastic neutron scattering measurements on a powder sample put an upper limit on the magnon gap $\Delta \leq 2$\,meV, and a sinusoidal-like dispersive signal at 5\,meV is reported at  scattering vectors, $\boldsymbol Q$, with short length~\cite{Choi12} but the powder experiment cannot attribute this dispersion to a direction in reciprocal space. 

We report single-crystal INS measurements aiming at mapping the low-energy spin-wave dispersion, in order to complement the high-energy data obtained by RIXS~\cite{Kim20} and in order to \pred{evaluate} the microscopic parameters of the extended $HK\Gamma\Gamma'$  Hamiltonian defined in Eq.~\ref{hamiltonian_Kitaev}, see Sec.~\ref{INS_LSWT}. 
\pred{We first describe our \nio \ experiments on the cold TAS {\it Thales} \cite{data-thales}.}
With the intention of resolving the energy gap $\Delta$ at the AFM zone center (0,\,1,\,0.5), we performed constant-$\boldsymbol Q$ scans on {\it Thales} that show the emergence of a low-energy magnon mode at 
$\Delta$\,=\,1.7(1)\,meV in the ordered phase at 1.5K. This finite-energy signal vanishes above $T_{\rm N}$, see Fig.~\ref{NIO_magnon_Thales}(a). Slightly above the transition, at 18\,K, one recognizes enhanced quasi-elastic critical scattering that becomes suppressed upon further heating to 40\,K. 

\pred{
Following the discussion in Ref.~\cite{Chaloupka15}, the strong anisotropy of the magnetic interaction modifies the overall aspect
of the magnon dispersion. Instead of transversal low-energy magnons appearing at the AFM zone centers of the Brillouin zone soft excitations are expected at the zone boundaries \pred{of the AFM zigzag structure}. 
Indeed, recent polarized INS studies on $\alpha$-RuCl$_3$ revealed that the low-energy magnon commonly detected at the AFM zone center emerges unexpectedly in the longitudinal channel, i.e.\ polarized along the static moment direction~\cite{Braden2025}. Therefore, this spin-wave excitation cannot be explained as a transversal magnon but results from zone-boundary modes that are folded to the AFM Bragg point due to the multi-domain arrangement inherent to the $C_3$ symmetry of the honeycomb \pred{layer, see introduction}. We thus emphasize that the signal and gap measured at the AFM zone center in \nio \ is not the common magnon anisotropy gap, but instead it is associated with the soft excitations reflecting
the non-condensed zigzag instabilities at the Brillouin zone boundaries (0.5,\,$\pm$0.5,\,0). 
}

\begin{figure*}[t!]
	\centering
	\includegraphics[width=1.95\columnwidth]{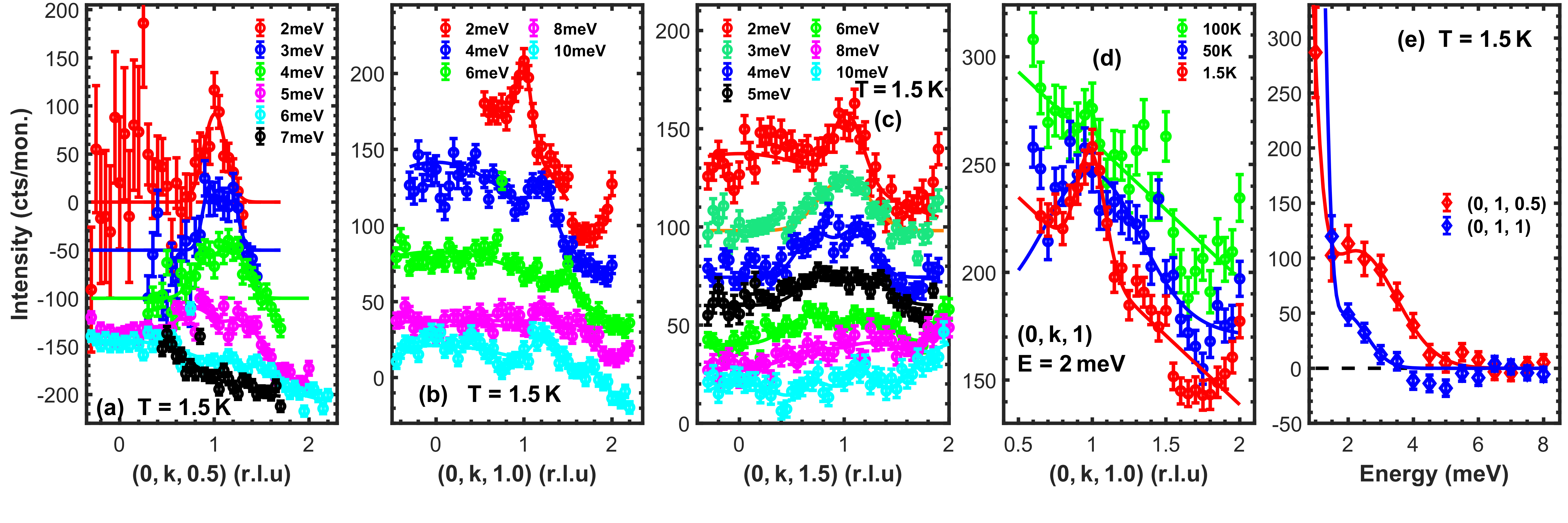}
	\caption{\label{Fig-IN8-1} Data obtained on $IN8(1)$ normalized to $2\cdot10^5$ monitor counts. \pred{Constant-$E$ scans across AFM zone centers at (0,\,1,\,0.5) (a), (0,\,1,\,1) (b) and (0,\,1,\,1.5) (c) at $T$ = 1.5\,K (some data are vertically shifted \cite{vertical_shifts}). In (a) background scans (collected after rotation of the sample by 45$^{\circ}$) have been subtracted from the data sets with $E$\,=\,2,\,3 and 4\,meV. 	Panel (d) compares the $E=2$\,meV data for different temperatures.	Panel (e) shows energy scans at the AFM zone centers  (0,\,1,\,0.5) and (0,\,1,\,1)	after subtraction of the background measured by rotating the sample by 50$^{\circ}$. }}
\end{figure*}

To investigate the out-of-plane dispersion of this low-energy excitation, similar constant-$\boldsymbol Q$ scans at the 2D AFM zone centers (0,\,1,\,$l$) with 0 $\le l \le$ 1 have been collected at $T=1.5$\,K, see Fig.~\ref{NIO_magnon_Thales}(b). Background data sets are collected by performing similar energy scans after rotating the sample by $\pm$45\,$^{\circ}$ outside of any magnetic signal, and are subtracted from the data. The original data sets without background subtraction and some additional scans collected at other $l$ values are shown in the Supplemental Material \cite{suppl-mat}. The magnon peak is  fitted with a single Gaussian function. 
The $l$ dependence of the energy gap, plotted in Fig.~\ref{NIO_magnon_Thales}(d), is insignificant compared to the total energy range of magnetic excitations, which documents the 2D nature of the magnetic excitations, and the absence of sizable interlayer coupling. The dashed-dotted line is a guide to the eye using a cosine function $a\cdot \left[\cos\left(2\pi k \right)+1\right]+\Delta$, where $a$\,=\,0.2\,meV and $\Delta$\,=\,1.7(1)\,meV.
%and with an out-of-plane spin-wave velocity $v_o=0.6$\,meV$\cdot$r.l.u.$^{-1}$.

We performed constant-$E$ scans across the AFM zone center (0,\,1,\,$\frac{1}{2}$) on {\it Thales} to study the in-plane magnon dispersion, see Fig.~\ref{NIO_magnon_Thales}(c). While the scan at $E$\,=\,1\,meV does not exhibit any peak, the magnon signal emerges at $E = 1.5$\,meV. At $E=2.25$\,meV, because of the increase in background, a similar scan has been collected after rotating the sample by $45^{\circ}$ where no magnetic signal is expected, and subtracted to the data set. For each data set below $E$\,=\,2\,meV, background has been modeled by a Gaussian function centered at $k$\,=\,0, and \pred{ the magnon peaks dispersing from $k=1$ are modeled with a single or two symmetric Gaussian functions of equal width, for $E \le 1.75$\,meV and $E > 1.75$\,meV, respectively.} 
The Gaussian peak positions reported in Fig.~\ref{NIO_magnon_Thales}(e) indicate the low-energy  magnon dispersion. \pred{The dashed-dotted line is a guide to the eye using a parabolic dispersion, modeled with $E=(\Delta^2+c\cdot(k-1)^2)^{0.5}$ ($\Delta$\,=\,1.7(1)\,meV and $c$\,=\,361\,meV$^2\cdot$r.l.u.$^{-2}$).} Note that because of the finite resolution, the constant-$E$ scan at $E=1.5$\,meV already senses the magnon dispersion and shows a magnetic signal in Fig.~\ref{NIO_magnon_Thales}(c), while the energy gap is found at $\Delta=1.7(1)$\,meV, see energy scan at $T$\,=\,1.5\,K in Fig.~\ref{NIO_magnon_Thales}(a).

\pred{The thermal TAS {\it IN8} permits studying the excitations at higher energy with higher neutron flux but  broader resolution.
At $T=1.5$\,K, we performed constant-energy scans along (0,\,$k$,\,$l$) with $l$ fixed. 
Neglecting all inter-layer coupling, these scans cover the 2D AFM Bragg positions (0,\,$1$,\,$l$), denoted $Y$ points, as well as the 2D FM positions (0,\,$0$,\,$l$) corresponding to the 2D $\Gamma$ points.} While the chosen $l$ values do not significantly impact the dispersion,
they modify the absolute length of the scattering vector, $\vert Q \vert$, and thus the scattering angle, which determines the background. Background issues turned out crucial for \pred{these} INS experiments due to the small
amount of sample and the weak signal associated with a rapidly decaying form factor for a large $5d$ magnetic ion.  
$\vert Q \vert$ modifies the intensity of magnetic scattering through the square of the magnetic form factor. By taking out the sample we verified that a large part of the background did not stem from the sample but from the air and cryostat-material scattering as well as from rapid neutrons. This generates
a background that for fixed energy transfer strongly increases toward low scattering angles or 
accordingly toward low $\vert Q \vert$ values. 
Furthermore, reaching large $E$ requires enhancing  $\vert Q \vert$ to close the scattering triangle, ${\boldsymbol k}_i={\boldsymbol Q}+{\boldsymbol k}_f$ (with ${\boldsymbol k}_{i,f}$ the incoming and outgoing neutron momenta).
In order to map the magnetic excitations that appear near (0,\,$1$,\,$l$), different $l$ values must be studied.
\pred{$l$\,=\,0.5 implies a large signal but also enhanced background, in particular when $\boldsymbol Q$ approaches
(0,\,$0$,\,0.5), and it is not possible to study high energy transfer.} 
In contrast scans at \pred{ $l$\,=\,1 and 1.5} suffer less from
the background variation but more from the reduced signal strength. 
The background conditions could be considerably improved in the second part $IN8(2)$ of experiments using a different cryostat and tighter diaphragms, therefore we always specify which conditions apply. 
  
\begin{figure}[ht!]
	\centering
	\includegraphics[width=0.93\columnwidth]{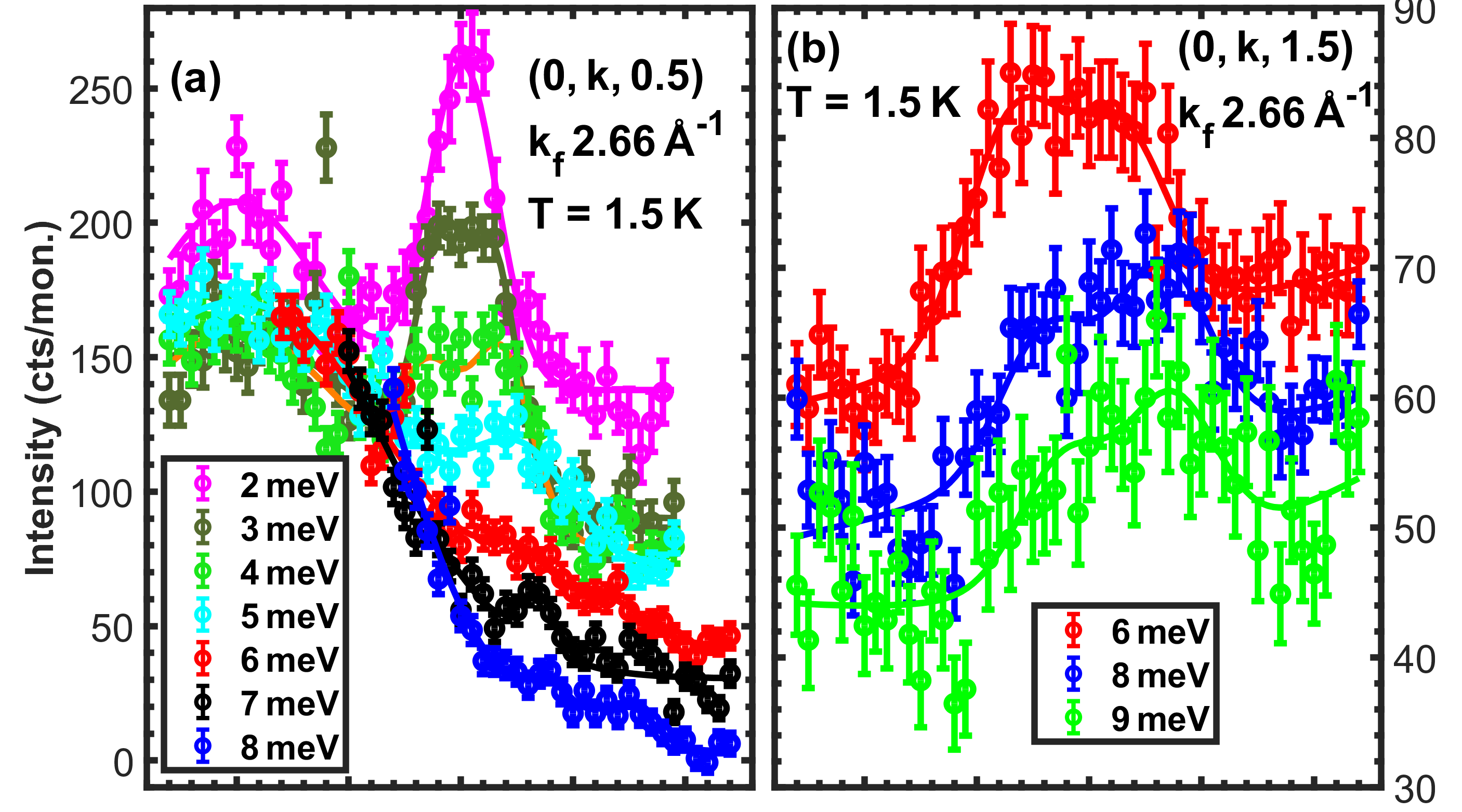}
	\includegraphics[width=0.93\columnwidth]{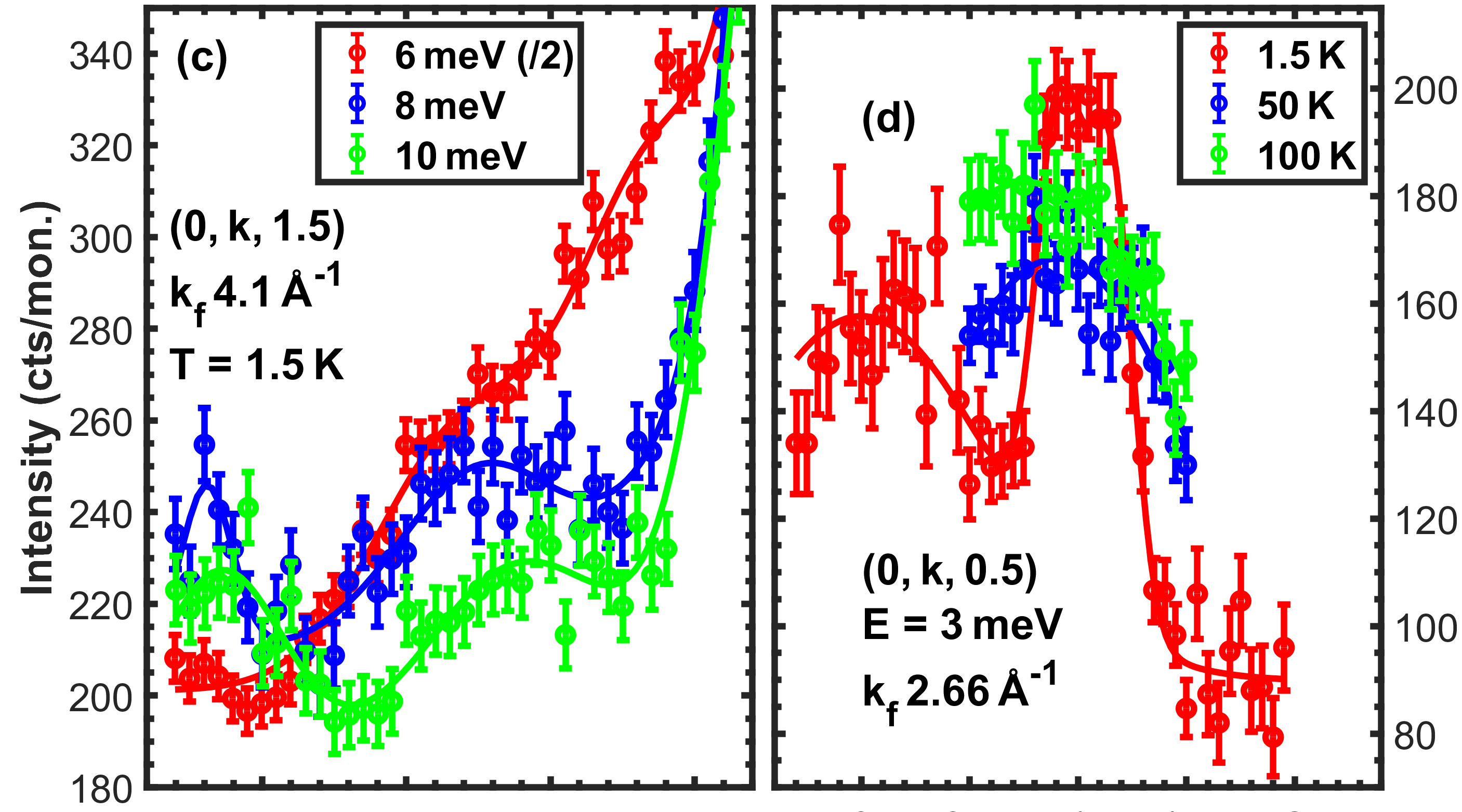}
	\includegraphics[width=0.93\columnwidth]{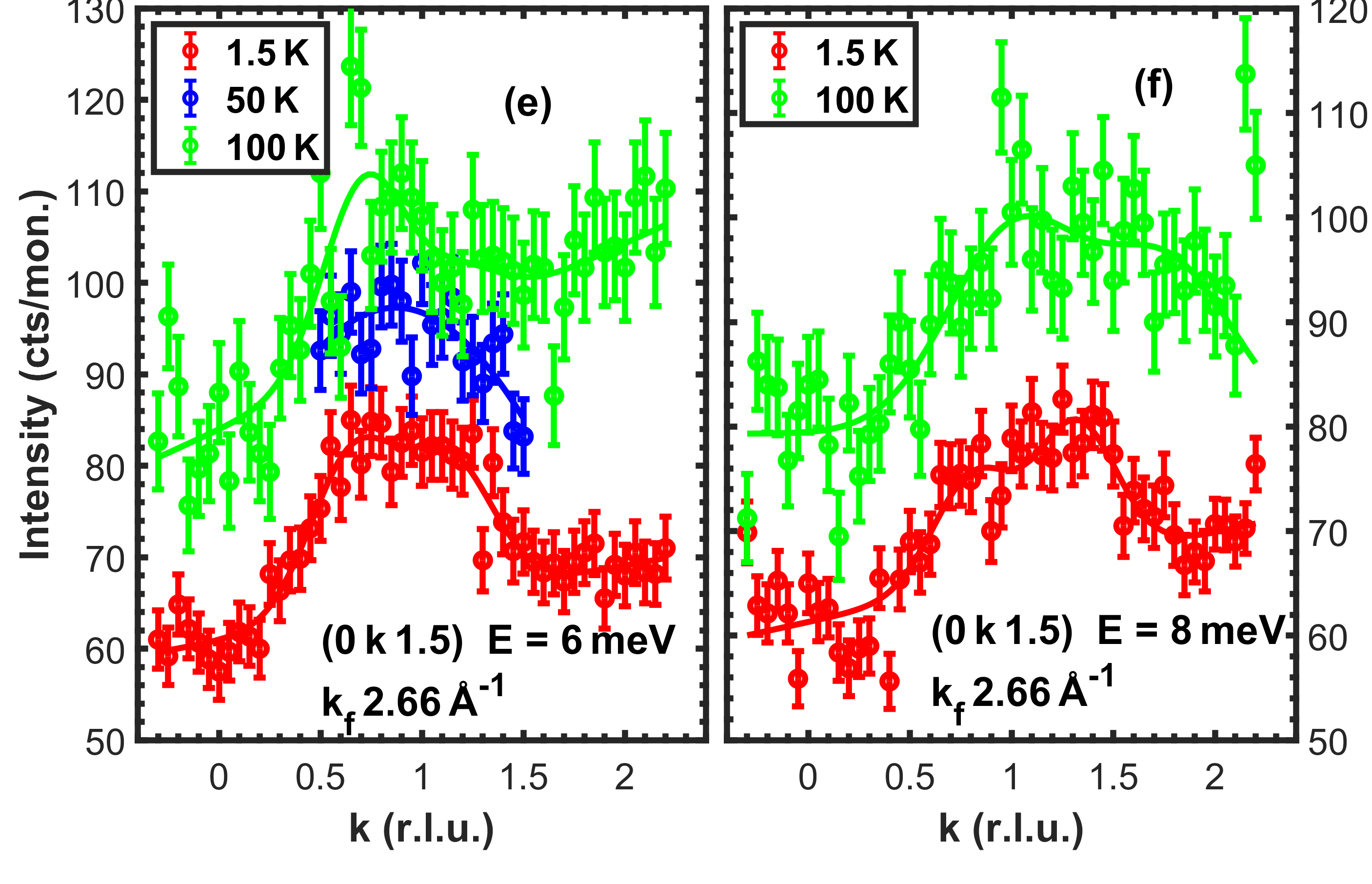}
	\caption{\label{Fig-IN8-2} 
	\pred{	Constant-$E$ scans along $k$ direction obtained on $IN8(2)$ normalized to $2\cdot10^5$ monitor counts:
		across the AFM zone centers (0,\,1,\,$l$) with $l=0.5$ (a) and $l=1.5$ (b), some data are shifted vertically \cite{vertical_shifts}. Panel (c) shows constant-$E$ scans across (0,\,1,\,1.5) measured with a larger final energy, $k_f$\,=\,4.1\,\AA$^{-1}$. Because of the high background, intensity of the 6\,meV scan has been divided by a factor 2. Panels (d-f) show constant-$E$ scans (with $k_f=2.662$\,\AA$^{-1}$) for
 		different temperatures. In all panels solid lines represent fits with Gaussian functions. }}
\end{figure}

Constant-$E$ scans at $l=0.5$, $1$ and $1.5$ across the 2D $\Gamma$ and $Y$ points are presented in Fig.~\ref{Fig-IN8-1} and Fig.~\ref{Fig-IN8-2} for the two experiments $IN8(1)$ and $IN8(2)$, respectively. 
Magnon modes can be resolved up to $E= 8, 10$, and 10\,meV for $l$\,=\,0.5, 1, and 1.5, respectively.  
The background is treated with a Gaussian function centered at $k$\,=\,0, and the magnon peaks dispersing from $k=1$ are modeled with two Gaussian functions of equal width. 
At $E=2$\,meV, only a single Gaussian peak is needed. 
\pred{These fits yield the dispersion data that are compared to the spin-wave calculations in Fig.\,6.
While the 2\,meV signal can be easily studied at \pred{ $l$\,=\,0.5,} higher energies require larger values, $l$\,=\,1 or 1.5, to suppress the steeply rising background toward small scattering angle.} 
Overall, the magnon disper\-sing away from the AFM zone center can be followed up to $\sim$10\,meV. 
Already the first inspection reveals that the magnon dispersion is rather unusual as the signal remains centered close to the $Y$ point at $k$\,=\,1 even for higher energies. 
The magnon $k$ positions obtained for different $l$ values in the two $IN8$ experiments and those obtained at $Thales$ are plotted together with the calculated magnon dispersion in Fig.~6(a), which illustrates the consistency of the three experiments.

Figure 4(e) presents energy scans for (0,\,1,\,0.5) and (0,\,1,\,1), which show the low-energy response
right at the 3D and 2D AFM zone centers, respectively, the \pred{stronger} signal at $l$\,=\,0.5 can be attributed to the larger form factor and to the 3D ordering. In the first $IN8(1)$ experiment we followed only the 2\,meV signal as a function of temperature, see Fig~4(d). The large and temperature dependent background hampers a quantitative analysis, but
correlations seem to persist well above $T_{\rm N}$. Figure~5(d-f) presents temperature dependent data for higher
energies and with the improved conditions of $IN8(2)$. \pred{The signal at 3\,meV survives} up to 100\,K but it becomes
significantly broadened indicating a shorter in-plane correlation length  for these fluctuations. The signals at 6 and 8\,meV also persist at 100\,K and there is no clear broadening, most likely because the signal is already
broad at low temperature due to the superposition of $k = 1\pm\xi$ contributions. The observation of dynamic magnetic correlations to about 6 times $T_{\rm N}$ underlines the 2D nature of magnetic \pred{interactions and supports coupling} parameters exceeding the simple  energy scale given by $T_{\rm N}$. Evidence for magnetic excitations at such temperatures was already deduced from RIXS studies \cite{Revelli20,Kim20}, but for much higher energies.

For \arc \, various INS experiments revealed coexisting strong FM fluctuations that at high temperatures even become truly dominant \cite{Banerjee17,do2017,Balz19,Braden2025}. Similar fluctuations must become visible
in our data covering not only the AFM scattering vectors (0,\,1,\,$l$) but also the FM ones (0,\,0,\,$l$).
The raw data, e.g. in Fig. 5(a), can suggest such a signal, but this enhanced intensity  can be entirely attributed to the background that increases for small scattering angle and thus for small $\vert Q \vert$.
In the first series of experiments $IN8(1)$ we determined the background by repeating scans with the sample rotated by $\pm$45\,$^\circ$. Fig. 4(a) presents the data after subtracting this background, which unambiguously shows that there is no comparably strong signal at the FM position. The data taken at 
\pred{$l$\,=\,1.5 and $E$\,=\,6,\,8 and 10\,meV} in Fig. 4(c) confirm this conclusion for higher energies. 
Also the second part $IN8(2)$ \pred{unambiguously shows the absence of such a FM signal for the entire energy range studied (between 2 and 9\,meV), see Fig.~5(b,c).}
Constant-$E$ scans at (0,\,$k$,\,1.5) have been repeated at $E=6, 8,$ and 10\,meV with a larger neutron final wave vector $k_f=4.1$\AA$^{-1}$, resulting in a larger resolution ellipsoid volume, thus strengthening the spectral weight of the hypothetical broad FM excitations. 
However, as shown in Fig.~5(c), there is no clear enhancement of the inelastic signal at $k=0$ underlining the absence of
a strong FM signal in \nio . 
Note that the spin wave modes around $k=1$ appear as a broad signal and not as a sharp peak because of the worsened energy resolution with $k_f=4.1$\,\AA$^{-1}$. 
The increase at larger $Q$ values results from crossing a low-lying optical phonon branch. Also the temperature dependent data in Fig.~\ref{Fig-IN8-2}(d-f) do not indicate an enhancement of scattering at the FM positions. We thus conclude
that the magnetic excitations in \nio \ fundamentally differ from those in \arc \ by the absence of FM scattering in the low energy range of AFM magnons. Therefore, the latter cannot be taken as a clear indication of FM Kitaev interaction.

\subsection{LSWT Simulation}
\label{INS_LSWT}

\pred{We analyze the magnon dispersion in \nio \ by LSWT applying one of the microscopic interaction models developed with the high-resolution RIXS data \cite{Kim20}. The spin-wave dispersion is computed with the package {\it spinW}~\cite{Toth15} implementing the pure 2D interaction model in the honeycomb lattice in $C2/m$ structure that is described by Equ.~(1). Additional calculations inserting other published interaction-parameter sets are presented and discussed in the Supplemental Material \cite{suppl-mat}. While the INS data are insufficient to refine the magnetic interaction parameters of such complex models they extend the RIXS results toward the low-energy range and the magnon gap, which is crucial to evaluate these models.  }

The Curie-Weiss temperatures obtained from susceptibility measurements, $\Theta_{\rm CW\,\parallel}=-176$\,K parallel to the layers and  $\Theta_{\rm CW\,\perp}=-40$\,K vertical, reflect the strong anisotropy of the material and indicate dominant AFM interactions~\cite{Singh10,Takagi19} but any quantitative interpretation must take the strong SOC into account. The zigzag  magnetic order can be explained with a simple isotropic Heisenberg Hamiltonian including AFM exchange parameters up to the 3rd-nearest neighbors, and the corresponding spin wave dispersion could reproduce the dispersive magnetic signal measured with powder INS~\cite{Choi12}. However, as shown in the Supplemental Material \cite{suppl-mat}, the computed spin wave dispersion cannot describe the experimental parabolic-like dispersion at low energy nor the peculiar high-energy dispersion deduced from the RIXS experiments \cite{Kim20}.

\begin{figure}[h!]
	\centering
	\includegraphics[width=0.99\columnwidth]{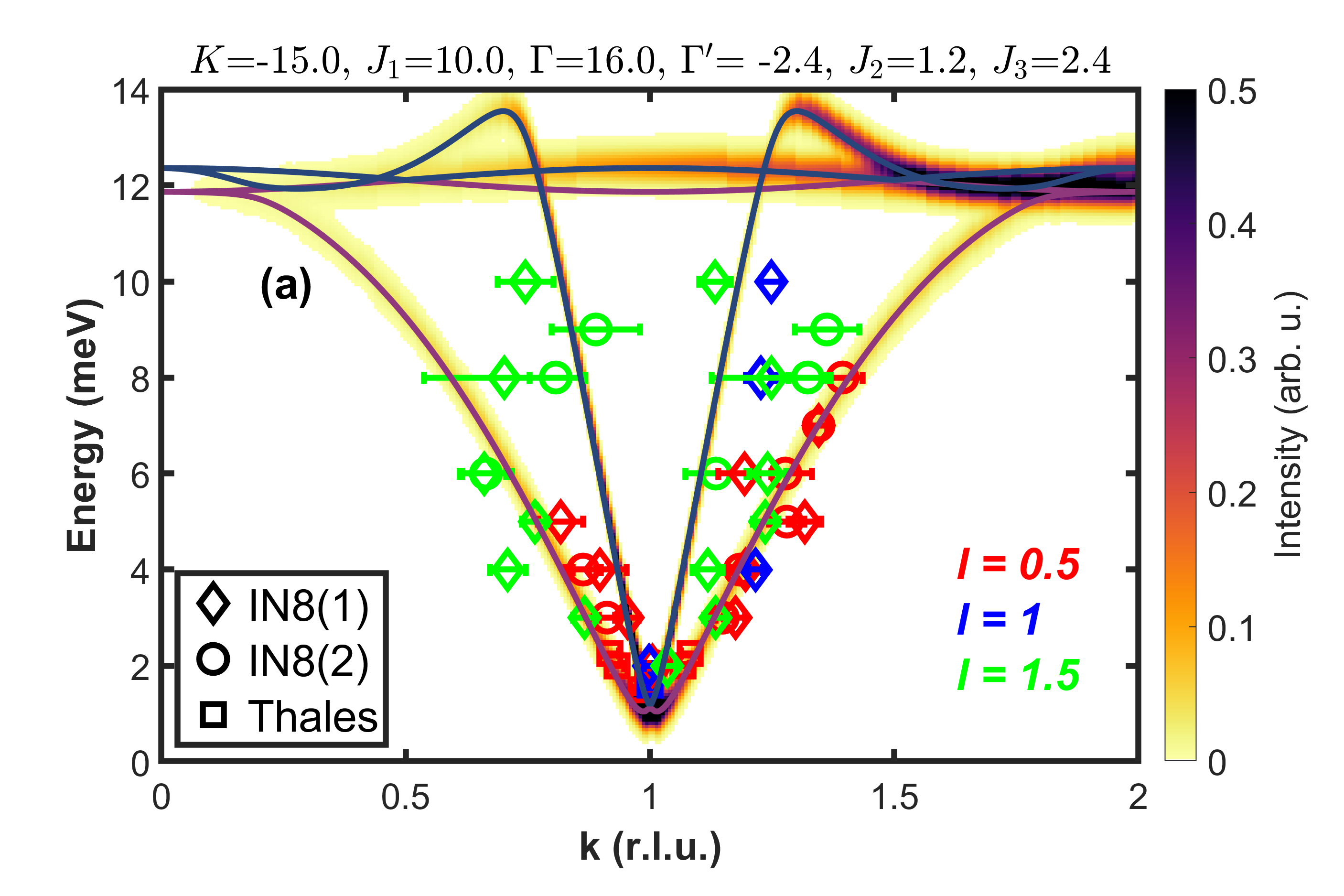}
	\includegraphics[width=0.99\columnwidth]{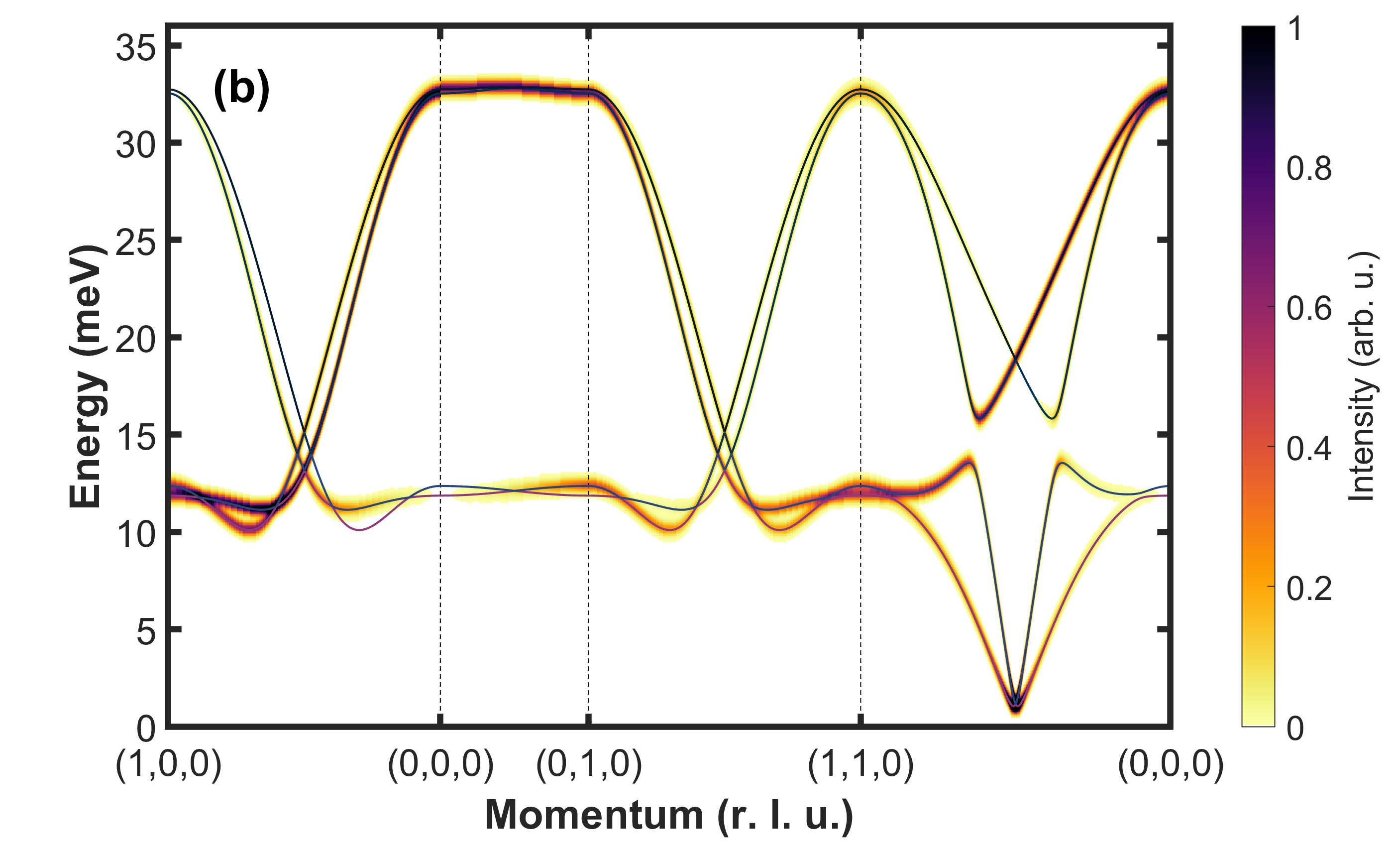}	
	\caption{\label{dispersion_full_1}  Calculation of the spin-wave dispersion with the interaction parameters of model A3 in Ref. \cite{Kim20} and comparison with our experimental data (a) and computation of the spin wave dispersion along high symmetry directions (b). 
		Only for panel (a) we superpose the contributions of the domains rotated by $\pm$120\,$^\circ$, which
		is mandatory to compare with the experimental data. The title line gives the interaction parameters in meV.}
\end{figure}

\pred{The pure Kitaev model yields a QSL ground state. In order to stabilize the zigzag magnetic ground state and to
describe the magnetic excitations, further couplings must be introduced leading to the commonly assumed $HK\Gamma\Gamma'$ model presented in Eq. (1). The complexity of this model is considerable hampering
an unambiguous determination of the parameters \cite{Kim20}. 
The orientation of the ordered moment can be used to restrict the parameter space, because the canting of the moment out of the layers
depends sensitively on the ratios between $J_1$, $K$ and $\Gamma$ \cite{Chaloupka16,Kim20,moeller2025}. However, the experimental precision of this canting angle \cite{Chun15}} is limited and the interpretation is affected by the coupling of spin and orbital contributions. For \arc \ polarized neutron diffraction finds a much lower canting compared to resonant x-ray diffraction, which can be at least partially attributed
to different sensitivities of the two experimental techniques \cite{Braden2025}. The RIXS data on \nio \ reveal 
magnetic excitations at the center of the Brillouin zone near 35\,meV which fixes the overall interaction
strength in \nio \ and which can discard many parameter sets, see Supplemental Material \cite{suppl-mat}.
\pred{Two models labeled A2 and A3 were proposed in Ref. \cite{Kim20} to correctly describe  
the high-energy part of the magnetic excitation dispersion determined by RIXS.} For these models little difference was found between LSWT and exact diagonalization calculations validating the LSWT approach. 
The two models though differ concerning the low-energy response, \pred{ in particular model A2 shows some instability around the $K$ point.}

Fig.~\ref{dispersion_full_1} presents the dispersion calculated with the parameter set of the $HK\Gamma\Gamma'$ model labeled A3 in Ref.~\cite{Kim20} \pred{that was implemented without any change}. 
Panel (a) directly compares the calculated dispersion with our INS results and gives the numerical values \pred{of the interaction terms} in the title line. 
Here, we sum over the calculated contributions from the three domains \pred{arising from the $C_3$ symmetry of the honeycomb layer}. 
Without any tuning of the parameters the agreement is good. The calculated dispersion \pred{well}
describes the low-energy magnon signal, that - we repeat - stems from the (0.5,\,$\pm$0.5,\,$l$) propagation vectors of
two rotated domains, \pred{but the experimental energy is slightly higher.}
The calculation also captures the low-energy dispersion of parabolic shape. 
The $k$ positions determined  at high energy that are close to $1$ can be attributed to an additional branch.
Since the model also describes the high-energy response observed in RIXS \cite{Kim20}, there is no reason
to further tune it. 
The full dispersion calculated for a single domain illustrates all the peculiarities of this model, see Fig.~6(b). 
There are no low-energy modes at the AFM zone center as discussed above and for \arc~\cite{Braden2025}, which is a direct consequence of the bond-directional anisotropy. 
The steepest dispersion emerges along [1,0,0] which is parallel to the \pred{ferromagnetically arranged} zigzag chains. 
Again this perfectly reflects the situation in \arc .
There is, however, an important difference between the \nio \ dispersion presented in Fig.~\ref{dispersion_full_1} and that in \arc .   \nio ~ does not exhibit a FM instability, as there are
no soft magnon excitations near the FM zone center, while \arc ~ exhibits at low temperature a strong mode at 2\,meV that furthermore is very sharp ~\cite{ran2017,Balz19,Banerjee18,Braden2025}.
Also at higher energy up to $\sim$9\,meV  \nio \ does not exhibit a FM signal
of comparable strength to the AFM magnon neither at 1.5\,K nor at higher temperature.
The presented interaction model for \nio \ perfectly captures this finding, see Fig. 6; such a FM mode is thus not
a direct consequence of the FM Kitaev parameter. Instead it must be attributed to the FM $J_1$ interaction of
\arc ~ that is AFM in \nio \ \cite{Kim20}, as it has been also deduced from theoretical estimations of the interaction parameters \cite{Liu22}.

\pred{The absence of the FM instability in our neutron data on \nio \ and the strength of such excitations in \arc \ \cite{Banerjee17,Banerjee18,do2017,Balz19,Braden2025} indicate the important qualitative difference of the magnetic interactions in these two materials. While in both compounds the FM Kitaev term is strong qualifying them unambiguously as Kitaev materials, there are other terms of comparable size, in particular $\Gamma$ and $J_1$. The sign change of $J_1$ between the two materials results in quite different nearest neighbor interaction matrices. For \arc \ the strongest eigenvalue still corresponds to the initial Kitaev term (using the set or parameters discussed in \cite{Braden2025}) but the two other Eigenvalues are of comparable size. With the \nio \ model discussed above, there is one dominating eigenvalue that, however, corresponds to the [011] direction for the $X$ bond. Both materials thus considerably deviate from the ideal Kitaev model, but in quite different ways. Whether it is still possible to observe signatures of the ideal Kitaev model in these materials remains an open challenge, but the deviations from Kitaev physics must be different in these two best characterized Kitaev candidate materials. }

\section{Conclusion}
 \label{conclusion}
 
We report on neutron scattering analyzes of \nio \ studying a sample consisting of 63 co-aligned crystals.
Magnetic superstructure reflections appear below $T_{\rm N}$ =\,15\,K and \pred{indicate magnetic order that is well defined perpendicular to the layers}. However, there is also quasistatic diffuse scattering indicating correlations with shorter 
coherence along $\bs c^*$ and magnetic stacking faults.

High-resolution INS scans indicate a small magnon gap of 1.7(1)\,meV, which closely resembles the finding in \arc . In analogy and in agreement with the LSWT calculation, this low-energy mode does not correspond to the usual transversal magnon at AFM zone center but stems from the zone boundary modes
of the two domains rotated by $\pm$120\,$^\circ$ against the domain that contributes to the magnetic
Bragg scattering. \pred{This mode shows almost no dispersion perpendicular to the
layers confirming the pronounced 2D nature of magnetic interactions in \nio .
By scanning with fixed energy transfer} along (0,\,1$\pm \xi$,\,$l$) we determine the low-energy part of the magnon dispersion, which complements the earlier RIXS studies at higher energies. There is a non-trivial dispersion
with some steepening of the dispersion towards higher energies. The comparison of the measured
dispersion with LSWT \pred{implementing interaction parameters} deduced from high-energy RIXS data~\cite{Kim20} yields satisfactory agreement
without any tuning of parameters corroborating this model. The parameter set supports strong
Kitaev interaction for \nio \ that is, however, not dominating.

The dispersion of magnetic excitations in \nio \ and in \arc \ \pred{resembles} each other concerning the low-energy mode stemming from the zone-boundary instability and the parabola-shaped dispersion, that are both captured by
the $HK\Gamma\Gamma'$ models for the two systems. There is however an important difference concerning the FM instability. \arc \ exhibits low-energy and very sharp modes at the 2D FM zone centers and this response even dominates the spectrum at higher temperatures. In \nio \ we do not find such a signal in spite of strong
efforts in the energy range 2 to 9\,meV where we can \pred{unambiguously} detect the magnon dispersion near the AFM positions. 
As it has been already deduced from theoretical analyses~\cite{Liu22}, \nio \ differs from its \arc \ sister compound concerning the absence of a FM instability in spite of both materials exhibiting a FM Kitaev term. The \pred{nearest-neighbor} Heisenberg term, however, is FM in \arc \ but AFM in \nio .

\begin{acknowledgments}
%	The supporting data and codes for this article are available from Zenodo \cite{zenodo}.
	AB and MB acknowledge funding from
	the Deutsche Forschungsgemeinschaft (DFG, German Research Foundation) under CRC 1238-277146847 (subproject
	B04). G.C. acknowledges the US National Science Foundation support via Grant No. DMR 2204811.
\end{acknowledgments}

\begin{center}
	{\bf DATA AVAILABILITY}
\end{center} 

The data that support the findings of this article are openly 
available \cite{zenodo}.

%\bibliography{NIO}	

\begin{thebibliography}{74}
\bibitem [{ver()}]{vertical_shifts}%
	\BibitemOpen
	\href@noop {} {}\bibinfo {note} {{ In Fig. 2(b) and (c), data collected at
			$T$\,=\,5,7,9,11,13,15,16\,K have been shifted by
			-40,\,-80,\,-120,\,-160,\,-200,\,-240,\,-280\,cts/s, respectively. In Fig.
			3(b) data at $l$\,=\,0.25,\,0.5,\,1 are shifted by -40,\,-80,\,-100 cts/mon.,
			respectively. In Fig. 3(c) data at $E$\,=\,1.5,\,1.75,\,2,\,2.25\,meV are
			shifted by -50,\,-100,\,-150,\,-350 cts/mon., respectively. In Fig. 4(a) the
			data collected at $E$=\,3,\,4,\,5,\,6,\,7\,meV have been shifted by
			-50,\,-100,\,-300,\,-300,\,-300\,cts/mon., respectively; in Fig. 4(b) data
			collected at $E$\,=2,\,4,\,6,\,8,\,10\,meV have been shifted by
			-50,\,-20,\,-40,\,-60,\,-90\,cts/mon., and in Fig. 4(c) data collected at
			$E$\,=2,\,4,\,5,\,6,\,8,\,10\,meV by
			-20,\,-10,\,-20,\,-30,\,-40,\,-60\,cts/mon., respectively. In Fig. 5(a) the
			$E$\,=\,6,\,7,\,8\,meV data are vertically shifted by -20,\,-40,\,-60
			cts/mon., and in Fig. 5(b) data at $E=8,\,9$\,meV are shifted by -10,\,-20
			cts/mon., respectively. }}\BibitemShut {NoStop}%
\end{thebibliography}
%Control: production of eprint (1) enabled
%

\end{document}